\documentclass[prc,preprint,showpacs,showkeys,nofootinbib]{revtex4}%
\usepackage{amssymb}
\usepackage{graphicx}
\usepackage{bm}
\usepackage{amsmath}
\usepackage{amsfonts}%
\begin{document}
\title{Intermediate energy Coulomb excitation as a probe of nuclear structure at
radioactive beam facilities}
\author{C.A.~Bertulani$^{1,3}$, A.E.~Stuchbery$^{1,2}$, T.J.~Mertzimekis$^{1}$ and
A.D.~Davies$^{1}$}
\affiliation{$^{1}$ National Superconducting Cyclotron Laboratory,
Michigan State University, East Lansing, Michigan, 48824, USA}
\affiliation{$^{2}$ Department of Nuclear Physics, Research School
Physical Sciences and Engineering, The Australian National University,
Canberra, ACT 0200, Australia }
\affiliation{$^{3}$ Department of Physics and Astronomy, Michigan State University, East
Lansing, Michigan, 48824, USA}

\begin{abstract}
The effects of retardation in the Coulomb excitation of radioactive nuclei in
intermediate energy collisions ($E_{lab}\sim100$ MeV/nucleon) are
investigated. We show that the excitation cross sections of low-lying states
in $^{11}$Be,$^{38,40,42}$S and $^{44,46}$Ar projectiles incident on gold and
lead targets are modified by as much as 20\% due to these effects. The angular
distributions of decaying gamma-rays are also appreciably modified.

\end{abstract}
\maketitle
\date{\today}



\section{\textbf{Introduction}}

The excitation of a nucleus by means of the electromagnetic interaction with
another nucleus is known as Coulomb excitation. Since the interaction is
proportional to the charge $Z$ of the nucleus, Coulomb excitation is
especially useful in the collision of heavy ions, with cross sections
proportional to $Z^{2}$. Pure Coulomb excitation is assured if the bombarding
energy is sufficiently below the Coulomb barrier. In this case the ions follow
a Rutherford trajectory and never come so close so that their nuclear matter
overlaps. This mechanism has been used for many years to study the
electromagnetic properties of low-lying nuclear states~\cite{AW75}.

The probability for Coulomb excitation of a nuclear state $\mid f\rangle$ from
an initial state $\mid i\rangle$ is large if the transition time $t_{fi}%
=\hbar/(E_{f}-E_{i})=1/\omega_{fi}$ is greater than the interaction time
$t_{coll}=a_{0}/v$, in a heavy ion collision with closest approach distance
$a_{0}$ and projectile velocity $v$. That is, the cross section for Coulomb
excitation is large if the \textit{adiabacity parameter} satisfies the
condition
\begin{equation}
\xi=\frac{t_{coll}}{t_{fi}}=\omega_{fi}\dfrac{a_{0}}{v}<1\ .\label{1.1}%
\end{equation}
This \textit{adiabatic cut-off} limits the possible excitation energies below
1-2 MeV in sub-barrier collisions.

A possible way to overcome this limitation, and to excite high-lying states,
would be the use of higher projectile energies. In this case, the closest
approach distance, at which the nuclei still interact only
electromagnetically, is of order of the sum of the nuclear radii,
$R=R_{P}+R_{T}$, where $P$ refers to the projectile and $T$ to the target. For
very high energies one has also to take into account the Lorentz contraction
of the interaction time by means of the Lorentz factor $\gamma=(1-v^{2}%
/c^{2})^{-1/2}$, with $c$ being the speed of light. For such collisions the
adiabacity condition, Eq.~(\ref{1.1}), becomes
\begin{equation}
\xi(R)=\frac{\omega_{fi}R}{\gamma v}<1\ .\label{1.3}%
\end{equation}
>From this relation one obtains that for bombarding energies around and above
100 MeV/nucleon, states with energy up to 10-20 MeV can be readily excited.
The experimental problem is to ensure that the collision impact parameter is
such that the nuclei do not overlap their matter distributions so that the
process consists of Coulomb excitation only. This has been achieved by a
careful filtering of the experimental events in terms of scattering angles,
multiplicity of particles, angular distributions, using light and heavy
targets, etc~\cite{HP76,Mot93,Iek93,Sch93,Rit93,Gl98,Dav98}.

The theory of Coulomb excitation in low-energy collisions is very well
understood~\cite{AW75}. It has been used and improved for over thirty years to
infer electromagnetic properties of nuclei and has also been tested in the
experiments to a high level of precision. A large number of small corrections
are now well known in the theory and are necessary in order to analyze
experiments on multiple excitation and reorientation effects.

The standard semiclassical theory of Coulomb excitation at low energies
assumes that the relative motion takes place on a classical Rutherford
trajectory, as long as the transition energy $E_{fi}=E_{f}-E_{i}$ is small
compared to the kinetic energy of the system. The cross section for exciting a
definite final state $\mid f\rangle$ from the initial state $\mid i\rangle$ is
then given by
\begin{equation}
\left(  \dfrac{d\sigma}{d\Omega}\right)  _{i\rightarrow f}=\left(
\dfrac{d\sigma}{d\Omega}\right)  _{Ruth}\cdot\;P_{i\rightarrow f}%
\ ,\label{1.4}%
\end{equation}
where $P_{i\rightarrow f}$ is the probability, evaluated in perturbation
theory, of the excitation of the target by the time-dependent electromagnetic
field of the projectile~\cite{AW75}.

In the case of relativistic heavy ion collisions pure Coulomb excitation may
be distinguished from the nuclear reactions by demanding extreme forward
scattering or avoiding the collisions in which violent reactions take
place~\cite{HP76}. The Coulomb excitation of relativistic heavy ions is thus
characterized by straight-line trajectories with impact parameter $b$ larger
than the sum of the radii of the two colliding nuclei. A detailed calculation
of relativistic electromagnetic excitation on this basis was performed by
Winther and Alder~\cite{WA79}. As in the non-relativistic case, they showed
how one can separate the contributions of the several electric ($E\lambda$)
and magnetic ($M\lambda$) multipolarities to the excitation. Later, it was
shown that a quantum theory for relativistic Coulomb excitation leads to minor
modifications of the semiclassical results~\cite{BB88}. In Ref.~\cite{BN93}
the connection between the semiclassical and quantum results was fully
clarified. More recently, a coupled-channels description of relativistic
Coulomb excitation was developed~\cite{BCG02}.

The semiclassical theory of Coulomb excitation for low energy collisions
accounts for the Rutherford bending of the trajectory, but relativistic
retardation effects are neglected, while in the theory of relativistic Coulomb
excitation recoil effects on the trajectory are neglected (one assumes
straight-line motion), but retardation is handled correctly. In fact, the
onset of retardation brings new important effects like the steady increase of
the excitation cross sections with increasing bombarding energy. In a heavy
ion collision around 100 MeV/nucleon the Lorentz factor $\gamma$ is about 1.1.
Since this factor enters in the excitation cross sections in many ways, as in
the adiabacity parameter, Eq.~(\ref{1.3}), one expects that some sizeable
($10-20\,\%$) modification of the theory of nonrelativistic Coulomb excitation
would occur. Also, recoil corrections are not negligible, and the relativistic
calculations based on a straight-line parameterization of the trajectory are
not completely appropriate to describe the excitation probabilities and cross sections.

These questions are very relevant, as Coulomb excitation has proven to be a
very useful tool in the investigation of rare isotopes in radioactive beam
facilities~\cite{Gl98}. Thus, it is appropriate to investigate the effects of
retardation and recoil corrections in Coulomb excitation at intermediate and
high energies. In this article we will assess these problems by using the
semiclassical approach of Ref.~\cite{AB89}. As we shall show in the next
sections, both retardation and recoil effects must be included for bombarding
energies in the range 30-200 MeV per nucleon.

This can be accomplished in a straight-forward way in the semiclassical
approach with a relativistic trajectory, appropriate for heavy ion collisions,
and the full expansion of the electromagnetic propagator~\cite{AB89}. In most
situations, the Coulomb excitation is a one-step process, which can be well
described in first-order perturbation theory. Exceptions occur for very
loosely bound nuclei, as for example the excitation of $^{11}$Li~\cite{Iek93},
or $^{8}$B~\cite{Mot93,Dav98}, in which case the electromagnetic transition
matrix elements are very large due to the small binding and consequent large
overlap with the continuum wavefunctions. Another exception is the excitation
of multiple giant resonances, due to the strong collective response of heavy
nuclei to the short electromagnetic pulse delivered in heavy ion collisions at
relativistic energies~\cite{Sch93,Rit93}.

\section{\textbf{Coulomb excitation from low to high energies}}

In the semiclassical theory of Coulomb excitation the nuclei are assumed to
follow classical trajectories and the excitation probabilities are calculated
in time-dependent perturbation theory. At low energies one uses Rutherford
trajectories~\cite{AW75} while at relativistic energies one uses
straight-lines for the relative motion~\cite{WA79,BB88}. In intermediate
energy collisions, where one wants to account for recoil and retardation
simultaneously, one should solve the general classical problem of the motion
of two relativistic charged particles. A detailed study of these effects has
been done in Refs.~\cite{AB89,AAB90}. In Ref.~\cite{AAB90} it was shown that
the Rutherford trajectory is modified as the retardation due to the
relativistic effects starts to set in. This already occurs for energies as low
as 10 MeV/nucleon. It was also shown that if we use the scattering plane
perpendicular to the $z$-axis, we may write the new Coulomb trajectory
parameterized by
\begin{equation}
x=a\,\left[  \cosh w+\epsilon\right]  ;\ \ \ \ \ \ \ \ \ y=a\,\sqrt
{\epsilon^{2}-1}\,\sinh w;\ \ \ \ \ \ \ \ \ \ \ z=0;\ \ \ \ \ \ \ t=\dfrac
{a}{v}\,\left[  w+\epsilon\,\sinh w\right]  \ ,\label{RT_1}%
\end{equation}
where $\epsilon=1/\sin(\Theta/2)$, with $\Theta$ being the deflection angle.
The impact parameter is related to the deflection angle by $b=a\cot\left(
\Theta/2\right)  $. The only difference from Eq.~(\ref{RT_1}) and the usual
parameterization of the Rutherford trajectory at non-relativistic energies is
the replacement of the \textit{half-distance of closest approach in a head-on
collision} $a_{0}={Z_{P}Z_{T}e^{2}/}m_{0}v^{2}$ by $a=a_{0}/\gamma$. This
simple modification agrees very well with numerical calculations based on the
Darwin Lagrangian and the next order correction to the relativistic
interaction of two charges~\cite{AAB90}.

Retardation also affects the dynamics of the Coulomb excitation mechanism and
needs to be included in collisions with energies around 100 MeV/nucleon and
higher. A detailed account of this has been given in Ref.~\cite{AB89}. The end
result is that the amplitude for Coulomb excitation of a target from the
initial state $\mid i\rangle$ to the final state $\mid f\rangle$ by a
projectile with charge $Z_{P}$ moving along a modified Rutherford trajectory
is given by
\begin{equation}
a_{fi}=\dfrac{Z_{P}e}{i\hbar}\sum_{\lambda\mu}\,\dfrac{4\pi}{2\lambda
+1}\,(-1)^{\mu}\,\left\{  S(E\lambda,\mu)\mathbf{\mathcal{M}}_{fi}%
(E\lambda,\,-\mu)+S(M\lambda,\mu)\mathbf{\mathcal{M}}_{fi}(M\lambda
,\,-\mu)\right\}  \ ,\label{aif}%
\end{equation}
where $\mathbf{\mathcal{M}}_{fi}(\pi\lambda,\mu)$ are the matrix elements for
electromagnetic transitions, defined as
\begin{equation}
\mathbf{\mathcal{M}}_{fi}(E\lambda,\mu)={\frac{(2\lambda+1)!!}{\kappa
^{\lambda+1}c(\lambda+1)}}\,\int\mathbf{j}_{fi}(\mathbf{r})\cdot
\mathbf{\nabla}\times\mathbf{L}\left\{  j_{\lambda}(\kappa r)\;Y_{\lambda\mu
}(\vartheta,\phi)\right\}  \,d^{3}r\ \label{MEl}%
\end{equation}
where $L=-i\mathbf{r}\times\mathbf{\nabla}$ and
\begin{equation}
\mathbf{\mathcal{M}}_{fi}(M\lambda,\mu)=-{\frac{i(2\lambda+1)!!}%
{\kappa^{\lambda}c(\lambda+1)}\ }\int\mathbf{j}_{fi}(\mathbf{r})\cdot
\mathbf{L}\left\{  j_{\lambda}(\kappa r)\;Y_{\lambda\mu}(\vartheta
,\phi)\right\}  \,d^{3}r\ ,\label{MMl}%
\end{equation}
with $\omega$ defined as the excitation frequency $\hbar\omega=E_{f}-E_{i}$
and $\kappa=\omega/c$. Using the Wigner-Eckart theorem
\begin{equation}
\mathbf{\mathcal{M}}_{fi}(\pi\lambda,-\mu)=\left(  -1\right)  ^{I_{f}-M_{f}%
}\ \left(
\begin{array}
[c]{ccc}%
I_{f} & \lambda & I_{i}\\
-M_{f} & \mu & M_{i}%
\end{array}
\right)  \ \left\langle I_{f}\left\Vert \mathbf{\mathcal{M}}(\pi
\lambda)\right\Vert I_{i}\right\rangle \ ,\label{WEck}%
\end{equation}
the geometric coefficients can be factorized in Eq.~(\ref{aif}).

The orbital integrals $S(\pi\lambda,\mu$) are given by (after performing a
translation of the integrand by $w\rightarrow w+i\left(  \pi/2\right)  $)
\begin{align}
S(E\lambda,\mu)  &  =\dfrac{\mathcal{C}_{\lambda\mu}}{va^{\lambda}}\;
I(E\lambda,\mu),\label{S}\\
S(M\lambda,\mu)  &  =-\dfrac{{\mathcal{C}_{\lambda+1,\mu}\,}}{\lambda
ca^{\lambda}}{\ [(2\lambda+1)/(2\lambda+3)]^{1/2}\,[(\lambda+1)^{2}-\mu
^{2}]^{1/2}\,\cot}\left(  {\vartheta/2}\right)  {\ \ }I(M\lambda
,\mu)\ ,\nonumber
\end{align}
with
\begin{equation}
\mathcal{C}_{\lambda\mu}=\left\{
\begin{array}
[c]{ccl}%
\sqrt{\dfrac{2\lambda+1}{4\pi}}\,\dfrac{\sqrt{(\lambda-\mu)!(\lambda+\mu)!}
}{(\lambda-\mu)!!(\lambda+\mu)!!}\;(-1)^{(\lambda+\mu)/2} & , &
\ \ \ \ \mathrm{for}\quad\lambda+\mu=\mathrm{even}\\
0 & , & \ \ \ \ \mathrm{for}\quad\lambda+\mu=\mathrm{odd}\,\ ,
\end{array}
\right. \label{Clm}%
\end{equation}
and\footnote{There is a misprint in the power of $v/c$ in Eqs. 3.11 and 3.15
of Ref.~\cite{AB89}.}
\begin{align}
I(E\lambda,\mu)  &  =-i({\frac{v\zeta}{c}})^{\lambda+1}\,{\frac{1}
{\lambda(2\lambda-1)!!}}\,e^{-\pi\zeta/2}\,\int_{-\infty}^{\infty
}dw\;e^{-\zeta\epsilon\cosh w}\,e^{i\zeta w}\nonumber\\
&  \times\dfrac{(\epsilon+i\sinh w-\sqrt{\epsilon^{2}-1}\cosh w)^{\mu}
}{(i\epsilon\sinh w+1)^{\mu-1}}\nonumber\\
&  \times\left[  (\lambda+1)\,h_{\lambda}-zh_{\lambda+1}-\dfrac{v}{c}
\epsilon\,\zeta\,\cosh w\cdot h_{\lambda}\right]  \ ,\label{IEl}%
\end{align}
and
\begin{align}
I(M\lambda,\mu)  &  =\frac{i(v\zeta/c)^{\lambda+1}}{(2\lambda-1)!!}
\ e^{-\pi\zeta/2}\,\int_{-\infty}^{\infty}dw\;e^{-\zeta\epsilon\,\cosh
w}e^{i\zeta w}\nonumber\\
&  \times\,\dfrac{(\epsilon+i\sinh w-\sqrt{\epsilon^{2}-1}\,\cosh w)^{\mu}
}{(i\epsilon\sinh w+1)^{\mu}}\ h_{\lambda}(z)\ .\label{IMl}%
\end{align}

In the above equations, all the first-order Hankel functions $h_{\lambda}$ are
functions of
\begin{equation}
z={\frac{v}{c}}\zeta\,(i\epsilon\sinh w+1)\,,\label{Psi}%
\end{equation}
with
\begin{equation}
\zeta=\dfrac{\omega a}{v}=\dfrac{\omega a_{o}}{\gamma v}{\ .}\label{eta}%
\end{equation}

The square modulus of Eq.~(\ref{aif}) gives the probability of exciting the
target from the initial state $\mid I_{i}M_{i}\rangle$ to the final state
$\mid I_{f}M_{f}\rangle$ in a collision with the center of mass scattering
angle $\vartheta$. If the orientation of the initial state is not specified,
the cross section for exciting the nuclear state of spin $I_{f}$ is
\begin{equation}
d\sigma_{i\rightarrow f}=\frac{a^{2}\epsilon^{4}}{4}\;\frac{1}{2I_{i}+1}
\;\sum_{M_{i},M_{f}}\mid a_{fi}\mid^{2}\,d\Omega\,,\label{cross_1}%
\end{equation}
where $a^{2}\epsilon^{4}d\Omega/4$ is the elastic (Rutherford) cross section.
Using the Wigner-Eckart theorem, Eq.~(\ref{WEck}), and the orthogonality
properties of the Clebsch-Gordan coefficients, gives
\begin{equation}
{\frac{d\sigma_{i\rightarrow f}}{d\Omega}}=\frac{4\pi^{2}Z_{P}^{2}e^{2}}
{\hbar^{2}}\;a^{2}\epsilon^{4}\;\sum_{\pi\lambda\mu}{\dfrac{B(\pi\lambda
,I_{i}\rightarrow I_{f})}{(2\lambda+1)^{3}}}\;\mid S(\pi\lambda,\mu)\mid
^{2}\ ,\label{cross_2}%
\end{equation}
where $\pi=E$ or $M$ stands for the electric or magnetic multipolarity, and
\begin{align}
B(\pi\lambda,I_{i}  &  \longrightarrow I_{f})=\dfrac{1}{2I_{i}+1}\;\sum
_{M_{i},M_{f}}\mid\mathcal{M}(\pi\lambda,\mu)\mid^{2}\ \nonumber\\
&  =\frac{1}{2I_{i}+1}\ \left\vert \left\langle I_{f}\left\Vert
\mathbf{\mathcal{M}}(\pi\lambda)\right\Vert I_{i}\right\rangle \right\vert
^{2}\ \label{reduced}%
\end{align}
is the reduced transition probability.

\section{Relativistic and non-relativistic limits}

The non-relativistic limit is readily obtained by using $v/c\rightarrow0$ in
the expressions in section II. In this case, $z\rightarrow0$ in Eq.~(\ref{Psi}%
), and
\begin{align}
h_{\lambda}  &  \rightarrow-i\left(  2\lambda-1\right)  !!\frac{1}
{z^{\lambda+1}}\nonumber\\
(\lambda+1)\,h_{\lambda}-zh_{\lambda+1}-\dfrac{v}{c}\epsilon\,\zeta\,\cosh
w\cdot h_{\lambda}  &  \rightarrow i\lambda\left(  2\lambda-1\right)
!!\frac{1}{z^{\lambda+1}}\ ,\label{approx1}%
\end{align}
which yields
\begin{align}
I(E\lambda,\mu)  &  =\,e^{-\pi\zeta/2}\,\int_{-\infty}^{\infty}dw\;e^{-\zeta
\epsilon\cosh w+i\zeta w}\dfrac{(\epsilon+i\sinh w-\sqrt{\epsilon^{2}-1}\cosh
w)^{\mu}}{(i\epsilon\sinh w+1)^{\lambda+\mu}},\label{Obtnr}\\
I(M\lambda,\mu)  &  =I(E\lambda+1,\mu)\ .
\end{align}

These are indeed the orbital integrals for non-relativistic Coulomb
excitation, as defined in Eq. (II-E.49) of Ref.~\cite{AW65}.

In the relativistic limit, $v/c\rightarrow1,$ $\zeta\rightarrow0$ in
Eq.~(\ref{eta}) and $\epsilon\simeq b/a\rightarrow\infty$. However, the
combination
\begin{equation}
\xi(b)=\zeta\epsilon=\frac{\omega_{fi}b}{\gamma v}%
\end{equation}
can remain finite.

The results for the orbital integrals can be expressed in closed analytical
forms. First we translate back the integrands in Eqs.~(\ref{IEl})
and~(\ref{IMl}) by $-i\left(  \pi/2\right)  $ to get
\begin{align}
I(E\lambda,\mu)  &  =-i({\frac{v\zeta}{c}})^{\lambda+1}\epsilon{\frac
{1}{\lambda(2\lambda-1)!!}}\,\int_{-\infty}^{\infty}dw\;e^{i\xi\sinh
w}\,\dfrac{(1+i\sinh w)^{\mu}}{(\cosh w)^{\mu-1}}\nonumber\\
&  \times\left[  (\lambda+1)\,h_{\lambda}-zh_{\lambda+1}+i\dfrac{v}{c}
\xi\sinh w\cdot h_{\lambda}\right]  \ ,
\end{align}
and
\begin{equation}
I(M\lambda,\mu)=\frac{i(v\zeta/c)^{\lambda+1}}{(2\lambda-1)!!}\ \,\int
_{-\infty}^{\infty}dw\;h_{\lambda}(z)\;e^{i\xi\,\sinh w}\dfrac{(1+i\sinh
w)^{\mu}}{(\cosh w)^{\mu}}\ ,
\end{equation}
where now $z={\dfrac{v}{c}}\xi\cosh w$, and we took the limit $\zeta
\rightarrow0$ and $\epsilon\rightarrow\infty.$ For the lowest multipolarities
these integrals can be obtained in terms of modified Bessel functions by
assuming the long-wavelength approximation, $\xi\left(  R\right)  \ll1$, valid
for almost all cases of practical interest. In this case, we can also use the
approximation of Eqs.~(\ref{approx1}). From Eq.~(\ref{RT_1}), in the
relativistic limit, $\sinh w=vt/b$ and $r=b\cosh w$. Thus, the integrals can
be rewritten as
\begin{align}
I(E\lambda,\mu)  &  =va^{\lambda}\,e^{-\pi\zeta/2}\int_{-\infty}^{\infty
}dt\;e^{i\xi vt/b}\,\dfrac{(b+ivt)^{\mu}}{(b^{2}+v^{2}t^{2})^{\left(
\lambda+\mu+1\right)  /2}}\left[  1-i\frac{vt}{\lambda b}\dfrac{v}{c}%
\xi\right]  \ ,\\
I(M\lambda,\mu)  &  =va^{\lambda+1}\ e^{-\pi\zeta/2}\,\int_{-\infty}^{\infty
}dt\;e^{i\xi vt/b}\,\dfrac{(b+ivt)^{\mu}}{(b^{2}+v^{2}t^{2})^{\left(
\lambda+\mu+2\right)  /2}}\ .
\end{align}

These integrals can be calculated analytically~\cite{GR65} to give
\begin{equation}
I(E\lambda,\mu)=F(\lambda,\mu,\xi)-\dfrac{v}{\lambda
c}\xi\frac{dF(\lambda ,\mu,\xi)}{d\xi},\ \ \ \ \ \ \ \ \ \ \
I(M\lambda,\mu)=F(\lambda+1,\mu,\xi) \label{geez1}
\end{equation}
where
\begin{equation}
F(\lambda,\mu,\xi)=2\left(  -1\right)
^{\frac{\lambda+\mu}{2}}\left( \frac{a}{b}\xi\right)
^{\lambda}\sum_{n=-\lambda}^{\lambda}\frac{1}{2^{n} }\left(
-1\right)  ^{n-\mu}\frac{P_{\lambda-n}^{\left(  n-\mu,\
n+\mu\right) }\left(  0\right)  }{P_{\lambda}^{\left(  -\mu,\
\mu\right)  }\left( 0\right)  }K_{n}\left(  \xi\right)  \ .
\label{geez2}
\end{equation}
In this equation $P_{n}^{\left(  \alpha, \beta\right)  }$ are the Jacobi
polynomials, and $K_{n}\left(  x\right)  $ are modified Bessel functions.
Since $\lambda+\mu=$even (odd) for electric (magnetic) excitations, we only
need to calculate the integrals for $\mu=\pm1$ for the E1 multipolarity,
$\mu=0,\pm2$ for the E2 multipolarity, and $\mu=0$ for the M1 multipolarity, respectively.

To leading order in $\xi$,
\begin{align}
I(E1,\pm1)  &  =\frac{2a}{b}\mathcal{I}\left(  \xi\right)
,\ \ \ \ \ I(M1,0)=I(E2,0)=\frac{2a^{2}}{b^{2}}\mathcal{I}\left(  \xi\right)
,\ \ \ \ \ \ \ I(E2,\pm2)=\frac{2a^{2}}{3b^{2}}\mathcal{I}\left(  \xi\right)
.\nonumber\\
\mathcal{I}\left(  \xi\right)   &  =\xi K_{1}\left(  \xi\right)  =\left\{
\begin{array}
[c]{lll}%
1,\ \ \ \ \  & \mathrm{for} & \ \ \ \ \ \xi\lesssim1\\
0,\ \ \ \ \  & \mathrm{for} & \ \ \ \ \ \xi\gtrsim1\ .
\end{array}
\right. \label{I_HG}%
\end{align}

When inserted in Eqs.~(\ref{S}) and~(\ref{cross_2}) the above results yield
the correct relativistic Coulomb excitation cross sections~\cite{WA79} in the
long wavelength approximation. Thus, we have shown explicitly that the
Equations~(\ref{RT_1})-(\ref{cross_2}) reproduce the non-relativistic and
relativistic Coulomb excitation expressions, as proved numerically in
Ref.~\cite{AB89}. We can now analyze the intermediate energy region
($E_{Lab}\sim100$ MeV/nucleon), where most experiments with radioactive beams
are being performed.

\section{Gamma-ray angular distributions}

As for the non-relativistic case~\cite{WB65,AW65}, the angular distributions
of gamma rays following the excitation depend on the frame of reference
considered. It is often more convenient to express the angular distribution of
the gamma rays in a coordinate system with the $z$-axis in the direction of
the incident beam. This amounts in doing a transformation of the excitation
amplitudes by means of the rotation functions $D_{m^{\prime}m}^{j}$. The final
result is identical to the Eqs.~(II.A.66-77) of Ref.~\cite{AW65}, with the
non-relativistic orbital integrals replaced by Eqs.~(\ref{IEl}) and~(\ref{IMl}%
), respectively. The angular distribution of the gamma rays emitted into solid
angle $\Omega_{\gamma}$, as a function of the scattering angle of the
projectile ($\Theta,\Phi$), is given by
\begin{equation}
W\left(  \Omega_{\gamma}\right)  =\sum_{k\kappa}a_{k\kappa}^{\lambda}\left(
\Theta,\Phi,\zeta\right)  A_{k}^{(\lambda)}Y_{k\kappa}\left(  \Omega_{\gamma
}\right)  \ .\label{stat0}%
\end{equation}
In our notation, the $z$-axis corresponds to the beam axis, and the
$a_{k\kappa}^{\lambda}\left(  \Theta,\Phi,\zeta\right)  $ are given by
\begin{equation}
a_{k\kappa}^{\lambda}\left(  \Theta,\Phi,\zeta\right)  =b_{k\kappa}^{\lambda
}/b_{00}^{\lambda}\ ,\label{stat00}%
\end{equation}
where, for electric excitations~\cite{AW65},
\begin{align}
b_{k\kappa}^{E\lambda}\left(  \Theta,\Phi,\zeta\right)   &  =-\frac{2}
{\sqrt{2k+1}}\left(
\begin{array}
[c]{ccc}%
\lambda & \lambda & k\\
1 & -1 & 0
\end{array}
\right)  ^{-1}\sum_{\mu\mu^{\prime}\kappa^{\prime}}\left(  -1\right)  ^{\mu
}\left(
\begin{array}
[c]{ccc}%
\lambda & \lambda & k\\
\mu & -\mu^{\prime} & \kappa^{\prime}%
\end{array}
\right) \nonumber\\
&  \times Y_{\lambda\mu}\left(  \frac{\pi}{2},0\right)  Y_{\lambda\mu^{\prime
}}\left(  \frac{\pi}{2},0\right)  \ I(E\lambda,\mu)I(E\lambda,\mu^{\prime
})D_{\kappa^{\prime}\kappa}^{k}\left(  \frac{\pi}{2}+\frac{\Theta}{2}
,\frac{\pi}{2},\Phi\right)  \ .\label{stat000}%
\end{align}
In Eq.~(\ref{stat0}) the coefficients $A_{k}^{(\lambda)}$ are given by
\begin{equation}
A_{k}^{(\lambda)}=F_{k}\left(  \lambda,I_{i},I_{f}\right)  \sum_{ll^{\prime}
}F_{k}\left(  l,l^{\prime},I_{g},I_{f}\right)  \Delta_{l}\Delta_{l^{\prime}
}\ ,
\end{equation}
where $\left\vert \Delta_{l}\right\vert ^{2}$ is the intensity (in sec$^{-1}$)
of the 2$^{l}$-pole radiation in the $\gamma$-transition from the excited
state $g$ to the state $f$. Explicitly, the $l$-pole conversion coefficient
$\Delta_{l}$ is given by
\begin{equation}
\Delta_{\pi l}=\left[  \frac{8\pi\left(  l+1\right)  }{l\left[  \left(
2l+1\right)  !!\right]  ^{2}}\frac{1}{\hbar}\left(  \frac{\omega}{c}\right)
^{2l+1}\right]  ^{1/2}\left(  2I_{f}+1\right)  ^{-1/2}\left\langle
I_{f}\left\Vert i^{s(l)}\mathcal{M}(\pi l)\right\Vert I_{g}\right\rangle
\ ,\label{Delt}%
\end{equation}
with $s(l)=l$ for electric $\left(  \pi=E\right)  $ and $s(l)=l+1$ for
magnetic $\left(  \pi=M\right)  $ transitions. The product $\Delta_{l}%
\Delta_{l^{\prime}}$ is always real since $\left(  -1\right)  ^{s(l)}=$ (the
parity). The coefficients $F_{k}\left(  l,l^{\prime},I_{g},I_{f}\right) $ are
geometrical factors defined by
\begin{align}
F_{k}\left(  l,l^{\prime},I_{g},I_{f}\right)   &  =\left(  -1\right)
^{I_{f}+I_{g}-1}\sqrt{\left(  2l+1\right)  \left(  2l^{\prime}+1\right)
\left(  2I_{f}+1\right)  \left(  2k+1\right)  }\nonumber\\
&  \times\left(
\begin{array}
[c]{ccc}%
l & l^{\prime} & k\\
1 & -1 & 0
\end{array}
\right)  \left\{
\begin{array}
[c]{ccc}%
l & l^{\prime} & k\\
I_{f} & I_{f} & I_{g}%
\end{array}
\right\}  \ ,\label{geom1}%
\end{align}
and
\begin{equation}
F_{k}\left(  l,I_{i},I_{f}\right)  =F_{k}\left(  l,l,I_{i},I_{f}\right)
.\label{geom2}%
\end{equation}

The normalization of the coefficients $a_{k\kappa}^{\lambda}\left( \Theta
,\Phi,\zeta\right) $ is such that $a_{00}^{\lambda}\left(  \Theta,\Phi
,\zeta\right)  =1.$ Only terms with even $k$ occur in Eq.~(\ref{stat0}).

The total angular distribution of the gamma rays, which integrates over all
scattering angles of the projectile, is given by
\begin{equation}
W\left(  \theta_{\gamma}\right)  =\sum_{k}a_{k}^{\lambda}\left(  \zeta\right)
A_{k}^{(\lambda)}P_{k}\left(  \cos\theta_{\gamma}\right)  \ ,\label{stat}%
\end{equation}
where the $z$-axis corresponds to the beam axis and the statistical tensors
are given by
\begin{equation}
a_{k}^{\lambda}\left(  \zeta\right)  =b_{k}^{\lambda}/b_{0}^{\lambda
}\ ,\label{stat2}%
\end{equation}
where (for electric excitations)~\cite{AW65},
\begin{align}
b_{k}^{E\lambda}\left(  \zeta\right)   &  =-\frac{2}{\sqrt{2k+1}}\left(
\begin{array}
[c]{ccc}%
\lambda & \lambda & k\\
1 & -1 & 0
\end{array}
\right)  ^{-1}\sum_{\mu\mu^{\prime}\kappa}\left(  -1\right)  ^{\mu}\left(
\begin{array}
[c]{ccc}%
\lambda & \lambda & k\\
\mu & -\mu^{\prime} & \kappa
\end{array}
\right) \nonumber\\
&  \times Y_{\lambda\mu}\left(  \frac{\pi}{2},0\right)  Y_{\lambda\mu^{\prime
}}\left(  \frac{\pi}{2},0\right)  \int_{\epsilon_{0}}^{\infty}d\epsilon
\ \epsilon\ I(E\lambda,\mu)I(E\lambda,\mu^{\prime})Y_{k\kappa}\left(
\frac{\pi}{2},\frac{\pi}{2}+\sin^{-1}\left(  1/\epsilon\right)  \right)
\ .\label{stat3}%
\end{align}
$\epsilon_{0}$ is the minimum value of the eccentricity, associated with the
maximum scattering angle $\Theta_{0}$ by $\epsilon_{0}=1/\sin\left(
\Theta_{0}/2\right)  $. One can show that the coefficients $b_{k}^{E\lambda}$
are real, even if the orbital integrals are not. Their imaginary parts cancel
out in the sum over $\mu\mu^{\prime}$.

For $M1$ excitations
\begin{align}
b_{k}^{M1}  &  =-\frac{8}{\sqrt{2k+1}}\left(
\begin{array}
[c]{ccc}%
1 & 1 & k\\
1 & -1 & 0
\end{array}
\right)  ^{-1}\left(
\begin{array}
[c]{ccc}%
1 & 1 & k\\
0 & 0 & 0
\end{array}
\right)  \left[  Y_{20}\left(  \frac{\pi}{2},0\right)  \right]  ^{2}%
\nonumber\\
&  \int_{\epsilon_{0}}^{\infty}d\epsilon\ \epsilon\ \left(  \epsilon
^{2}-1\right)  \ I(M1,0)I(M1,0)Y_{k0}\left(  \frac{\pi}{2},\frac{\pi}{2}
+\sin^{-1}\left(  1/\epsilon\right)  \right)  .\label{stat32}%
\end{align}

The normalization of the coefficients $a_{k}^{\lambda}\left( \zeta\right) $ is
such that $a_{0}^{\lambda}\left(  \zeta\right) =1.$ Again, only terms with
even $a_{k}^{\lambda}\left( \zeta\right) $ occur in Eq.~(\ref{stat}).

In the case of $M1$ excitations Eq.~(\ref{stat32}) contains only $\kappa
=\mu=\mu^{\prime}=0$ and one gets $a_{2}^{M1}\left(  \zeta\right) =1,$
independent of $\zeta$. Since for small $\zeta$ the magnitude of
$a_{k}^{\lambda}\left(  \zeta\right) $ decreases appreciably with $k$ we will
only consider gamma ray emission after excitation through electric multipoles,
in particular, the dependence of $a_{2}^{E1}$ and $a_{2}^{E2}$ on $\zeta$.
This dependence is very weak at energies $E_{lab}\gtrsim100$ MeV/nucleon. In
that case, one can use the approximate relations, Eq.~(\ref{I_HG}), for
excitation energies such that $\xi\ll1$. This condition is met for reactions
with neutron-rich or proton-rich nuclei where the excitation energies involved
are of the order of $E_{x}\sim1$ MeV. It is then straightforward to show that
\begin{equation}
a_{2}^{E1}=1,\ \ \ \ \ a_{2}^{E2}=-2,\ \ \ \ \ \ \ \text{\ and} \ \ \ \ a_{4}%
^{E2}=-0.25.\label{ae1ae2}%
\end{equation}

We thus come to the important conclusion that in high energy collisions and
low excitation energies, $E_{x}\sim1$ MeV, the angular distribution of
gamma-rays from decays after Coulomb excitation does not depend on the
parameters $\zeta$ or $\xi$.

Although the cross sections for $M1$ and $E2$ excitations do not contain
interference terms, the $\gamma$-decay of the excited state can contain an
interference term with mixed $E2+M1$ multipolaritites. The angular
distribution of the gamma rays from the deexcitation of these states is given
by~\cite{AW65}
\begin{equation}
W^{E2,M1}\left(  \theta_{\gamma}\right)  =2\sqrt{\sigma_{M1}}\sqrt{\sigma
_{E2}}\sum_{k}a_{k}^{E2,M1}\left(  \zeta\right)  F_{k}\left(  1,2,I_{i}
,I_{f}\right)  \sum_{ll^{\prime}}\Delta_{l}\Delta_{l^{\prime}}F_{k}\left(
l,l^{\prime},I_{g},I_{f}\right)  P_{k}\left(  \cos\theta_{\gamma}\right)
\ ,\label{gyi}%
\end{equation}
where $\sigma_{M1}$ ($\sigma_{E2}$) is the total magnetic dipole (electric
quadrupole) excitation cross section and where the sign of the square root is
the same as the sign of the reduced matrix element $\left\langle i\left\Vert
\mathcal{M}\left(  M1\right)  \right\Vert f\right\rangle $ ($\left\langle
i\left\Vert \mathcal{M}\left(  E2\right)  \right\Vert f\right\rangle $). These
latter are the same as those occurring in the radiative decay
$f\longrightarrow i$. The $a_{k}^{E2,M1}$ coefficients in Eq.~(\ref{gyi}) are
given by
\begin{equation}
a_{k}^{E2,M1}=b_{k}^{E2,M1}/\sqrt{b_{0}^{E2}}\sqrt{b_{0}^{M1}}\ ,\label{gyi2}%
\end{equation}
where $b_{k}^{\lambda}$ is given by Eq.~(\ref{stat3}) and
\begin{align}
b_{k}^{E2,M1}  &  =-\frac{4}{\sqrt{2k+1}}\left(
\begin{array}
[c]{ccc}%
2 & 1 & k\\
1 & -1 & 0
\end{array}
\right)  ^{-1}\sum_{\mu}\left(  -1\right)  ^{\mu}\left(
\begin{array}
[c]{ccc}%
2 & 1 & k\\
\mu & 0 & \mu
\end{array}
\right) \nonumber\\
&  \times Y_{2\mu}\left(  \frac{\pi}{2},0\right)  Y_{20}\left(  \frac{\pi}
{2},0\right)  \int_{\epsilon_{0}}^{\infty}d\epsilon\ \epsilon\ \sqrt
{\epsilon^{2}-1}\ I(E2,\mu)I(M1,0)Y_{k\mu}\left(  \frac{\pi}{2},\frac{\pi}
{2}+\sin^{-1}\left(  1/\epsilon\right)  \right)  \ .\label{gyi3}%
\end{align}

Using the relations in Eq.~(\ref{I_HG}) and performing the summation above it
is straightforward to show that for $\xi(R)\ll1$
\[
a_{k}^{E2,M1}=0,\ \ \ \ \mathrm{for\ any}\ k\ .
\]
Thus, in high-energy collisions, there is no interference term from mixed
E2-M1 excitations in the angular distribution of emitted gamma rays.

The form of the expressions for the angular distribution used here has
followed that of Chapter 11 of Ref.~\cite{AW65}. Experimenters usually find it
more convenient to write the angular distribution in a slightly different form
which separates the statistical tensors that describe the orientation of the
state due to the excitation process from the geometrical factors associated
with the $\gamma$-ray decay and gives the geometrical factors the same form as
occurs in the formulation of $\gamma$-$\gamma$ correlations (cf.
Ref.~\cite{AW65} page 311; see also Refs.~\cite{stu02,stu03}). The general
expression for the $\gamma$-ray decay into solid angle $\Omega_{\gamma}$ after
projectile scattering to the angle $(\Theta, \Phi)$, where the transition
takes place between the Coulomb-excited state $f$ and a lower state $g$ (see
Eq.~(\ref{stat0})) becomes:
\begin{equation}
W\left(  \Omega_{\gamma}\right)  =\sum_{k\kappa}\alpha_{k\kappa}^{\lambda
}\left(  \Theta,\Phi,\zeta\right)  A_{k}(\delta_{\gamma} ll^{\prime}I_{g}%
I_{f}) Q_{k}(E_{\gamma})Y_{k\kappa}\left(  \Omega_{\gamma}\right)
\ ,\label{ADgen}%
\end{equation}
where the $A_{k}(\delta_{\gamma}ll^{\prime}I_{g}I_{f})$ coefficients are
related to the so-called $F$-coefficient (Eq.~(\ref{geom1})) for the $\gamma
$-ray transition between the states $I_{f}$ and $I_{g}$ with mixed
multipolarities $l$ and $l^{\prime}$ and mixing ratio $\delta_{\gamma}%
$~\cite{kra73,AW65} by the expression
\begin{equation}
A_{k}(\delta_{\gamma} ll^{\prime}I_{f}I_{g})=[F_{k}(llI_{f}I_{g}%
)+2\delta_{\gamma} F_{k}(ll^{\prime}I_{f}I_{g})+ \delta_{\gamma}^{2}%
F_{k}(l^{\prime}l^{\prime}I_{f} I_{g})]/(1+\delta_{\gamma}^{2}).\label{eq:A-F}%
\end{equation}
Note that for $k=0$ we have $A_{0}=F_{0}=1$, and due to the normalization
used, the matrix elements, Eq.~(\ref{Delt}), are not needed. In most
situations one is interested in the possible mixing of $E2$ and $M1$
multipolarities in the decay of $f\longrightarrow g$. Thus, one only needs the
$E2/M1$ mixing ratio. The quantity $Q_{k}(E_{\gamma})$ is the solid-angle
attenuation coefficient which takes account of the finite solid-angle opening
of the $\gamma$-ray detector~\cite{yates}. It follows that
\begin{equation}
\alpha_{k\kappa}^{\lambda}\left(  \Theta,\Phi,\zeta\right)  = a_{k\kappa
}^{\lambda}\left(  \Theta,\Phi,\zeta\right)  F_{k}(\lambda\lambda I_{i}I_{f})
.
\end{equation}

In the case where the particle scatters into an annular counter about the beam
direction (i.e. the $z$-axis), or for angular distributions where the
scattered particle is not detected at all (Eq.~(\ref{stat})), only $\kappa= 0$
terms survive. Usually the coefficients are normalized so that
\begin{equation}
W(\theta_{\gamma})=1+\sum_{k=2,4}B_{k}^{\lambda}\left(  \zeta\right)
A_{k}(\delta_{\gamma} ll^{\prime}I_{g}I_{f}) Q_{k}(E_{\gamma})P_{k}\left(
\cos\theta_{\gamma}\right)  .\label{eq:ADNorm}%
\end{equation}
The alignment of the initial state is now specified by the statistical tensor
$B_{k}^{\lambda}$ which is related to the statistical tensors introduced above
by
\begin{equation}
B_{k}^{\lambda} = \sqrt{2k +1} \; \alpha^{\lambda}_{k 0}/ \alpha^{\lambda}_{0
0},
\end{equation}
when the particle is detected in an annular counter, and
\begin{equation}
B_{k}^{\lambda} = a_{k}^{\lambda}\left( \zeta\right)  F_{k}(\lambda\lambda
I_{i}I_{f}) ,\label{Bkl}%
\end{equation}
when the particle is not detected at all (or detected in such a way as to
include all kinematically allowed scattering angles).

\section{Numerical results}

In Table~\ref{tab:Table_1} we show the numerical results for the orbital
integral $I(E2,\mu)$ for a deflection angle of 10$^{0}$ and for $\mu=2,0,-2$.
The calculations have been done using the code COULINT~\cite{Ber03}. The
results for $\gamma=1 $ agree within 1/1000 with the numerical values obtained
in Ref.~\cite{AW56}, also reprinted in Table II.12 of Ref.~\cite{AW65}. One
observes that the results of the integrals for $\gamma=1.1,$ corresponding to
a laboratory energy of about 100 MeV/nucleon, differ substantially from the
results for $\gamma=1$ (non-relativistic), specially for large values of
$\zeta$. For a fixed scattering angle $\zeta$ increases with the excitation
energy. Thus, one expects that the relativistic corrections are greater as the
excitation energy increases.

\begin{figure}[t]
\begin{center}
\includegraphics[
height=3.1747in,
width=3.192in
]{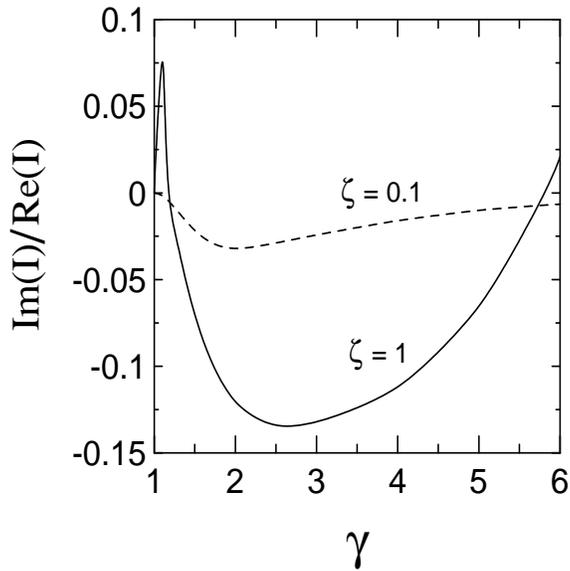}
\end{center}
\caption{The ratio of the imaginary to real parts of the orbital integral
$I(E2,2)$ is shown for $\varsigma=0.1$ (dashed curve) and $\zeta=1$ (solid
curve) as a function of $\gamma$.}%
\label{f1}%
\end{figure}

For $\gamma=1$ the imaginary part of the orbital integrals vanishes. But as
$\zeta$ and $\gamma$ increase the imaginary part becomes important. This is
shown in Fig.~\ref{f1} where the ratio of the imaginary to real parts of the
orbital integral $I(E2,2)$ is shown for $\zeta=0.1$ (dashed curve) and
$\zeta=1$ (solid curve) as a function of $\gamma$.

\begin{figure}[t]
\begin{center}
\includegraphics[
height=3.9539in,
width=3.2681in
]{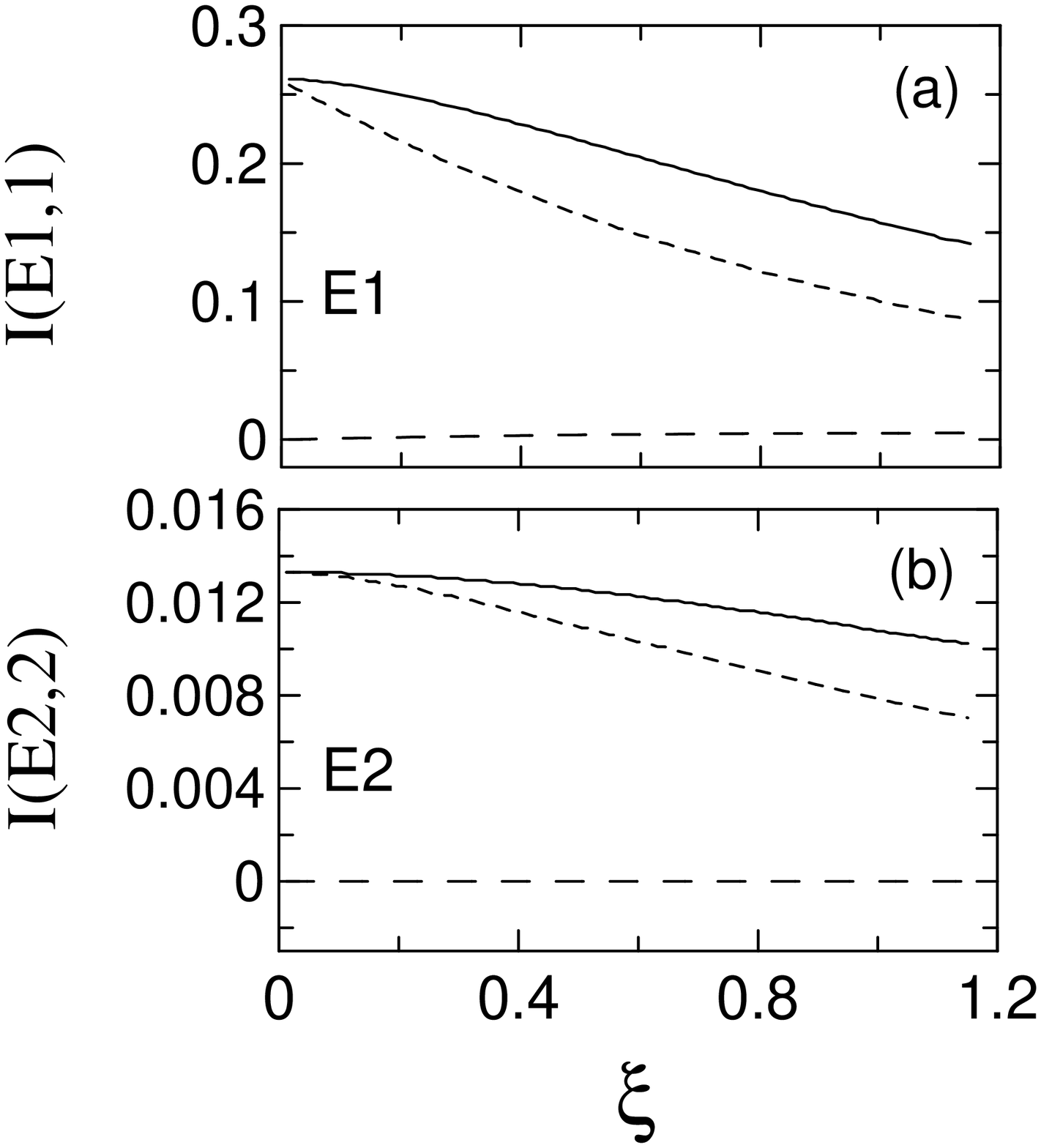}
\end{center}
\caption{ Upper figure: Real (solid lines) and imaginary part (long-dashed
lines) of the orbital integral $I(E1,1)$ for $\gamma=1.1$ ($E_{\mathrm{lab}%
}\simeq100$ MeV/nucleon)$.$ The approximation of Eq.~(\ref{I_HG}) is shown by
the dashed line. Lower figure: Same plot, but for the orbital integral
$I(E2,2)$. }%
\label{f2}%
\end{figure}

Except for the very low energies such that $a_{0}$ attains a large value, and
for the very large excitation energies $\hbar\omega$, the parameter $\zeta$ is
much smaller than unity. Also, at intermediate energies the scattering angle
is limited to very forward scattering. It is useful to compare the orbital
integrals with their limiting expressions given by Eq.~(\ref{I_HG}), i.e. the
relativistic limit to leading order in $\xi$. This is shown in Fig.~\ref{f2}%
(a) where the real (solid lines) and imaginary parts (long-dashed lines) of
the orbital integral $I(E1,1)$ are compared to the approximation of
Eq.~(\ref{I_HG}) (dashed line) for $\gamma=1.1$ ($E_{\mathrm{lab}}\simeq100$
MeV/nucleon). Figure~\ref{f2}(b) shows the same results, but for the orbital
integral $I(E2,2)$. The comparison is made in terms of the variable
$\xi=\epsilon\zeta$\, which is the appropriate variable for high energy
collisions. Only for $\xi\ll1$ do the expressions in the relativistic limit
reproduce the correct behavior of the orbital integrals. Although the
imaginary parts of the orbital integrals are small, the real parts show
substantial deviations from the approximations of Eq.~(\ref{I_HG}) at
intermediate energies ($E_{\mathrm{lab}}\simeq100$ MeV/nucleon).

\begin{table}[th]
\begin{center}%
\begin{tabular}
[c]{|l|l|l|l|}\hline
$\zeta$ & $\ \ \ \ \ \ \lambda,\mu=2,2$ & $\ \ \ \ \ \ \ \ \lambda,\mu=2,0$ &
$\ \ \ \ \ \ \ \lambda,\mu=2,-2$\\\hline
0.0 & 5.064(-3) [5.052(-3)] & 1.332(-2) [1.332(-2)] & 5.064(-3)
[5.073(-3)]\\\hline
0.1 & 8.675(-4) [1.105(-3)] & 6.505(-3) [7.621(-3)] & 1.195(-2)
[1.205(-2)]\\\hline
0.2 & 1.895(-4) [2.897(-4)] & 2.280(-3) [3.276(-3)] & 7.765(-3)
[9.487(-3)]\\\hline
0.3 & 4.425(-5) [7.849(-5)] & 7.311(-4) [1.296(-3)] & 2.245(-3)
[5.470(-3)]\\\hline
0.4 & 1.069(-5) [2.122(-5)] & 2.245(-4) [4.920(-4)] & 1.468(-3)
[2.728(-3)]\\\hline
0.5 & 2.637(-6) [5.599(-6)] & 6.716(-5) [1.821(-4)] & 5.442(-4)
[1.251(-3)]\\\hline
0.6 & 6.598(-7) [1.402(-6)] & 1.975(-5) [6.627(-5)] & 1.908(-4)
[5.431(-4)]\\\hline
0.7 & 1.668(-7) [3.169(-7)] & 5.739(-6) [2.383(-5)] & 6.438(-5)
[2.268(-4)]\\\hline

0.8 & 4.250(-8) [5.580(-8)] & 1.652(-6) [8.492(-6)] & 2.110(-5)
[9.201(-5)]\\\hline
0.9 & 1.090(-8) [1.906(-9)] & 4.724(-7) [3.004(-6)] & 6.765(-6)
[3.650(-5)]\\\hline
1.0 & 2.807(-9) [-5.003(-9)] & 1.343(-7) [1.057(-6)] & 2.131(-6)
[1.422(-5)]\\\hline
1.2 & 1.886(-10) [-1.845(-9)] & 1.071(-8) [1.291(-7)] & 2.031(-7)
[2.074(-6)]\\\hline
1.4 & 1.282(-11) [-3.551(-10)] & 8.426(-10) [1.556(-8)] & 1.858(-8)
[2.902(-7)]\\\hline
1.6 & 8.788(-13) [-5.699(-11)] & 6.566(-11) [1.857(-9)] & 1.651(-9)
[3.941(-8)]\\\hline
1.8 & 6.068(-14) [-8.381(-12)] & 5.078(-12) [2.200(-10)] & 1.433(-10)
[5.227(-9)]\\\hline
2.0 & 4.213(-15) [-1.170(-12)] & 3.904(-13) [2.591(-11)] & 1.222(-11)
[6.809(-10)]\\\hline
4.0 & 1.294(-26) [-1.211(-21)] & 2.362(-24) [1.111(-20)] & 1.464(-22)
[5.663(-19)]\\\hline
\end{tabular}
\end{center}
\caption{ The classical orbital integrals for E2 Coulomb excitation. The Table
lists the values of the classical orbital integrals $I(E2,\mu)$ for a
deflection angle of 10$^{\circ}$ and for $\mu=2,0,-2$. These entries are given
in the form of a number followed by the power of ten which it should be
multiplied. The value outside (inside) the brackets are for $\gamma=1$
($\gamma=1.1$).}%
\label{tab:Table_1}%
\end{table}

We now apply the formalism to specific cases. We study the effects
of relativistic corrections in the collision of the radioactive
nuclei $^{38,40,42}$S and $^{44,46}$Ar on gold targets. These
reactions have been studied at $E_{lab}\sim40$ MeV/nucleon at the
MSU facility~\cite{Sch96}. In the following calculations the
conditions may be such that there will be contributions from
nuclear excitation, but these will be neglected as we only are
interested in the relativistic effects in Coulomb excitation at
intermediate energy collisions. In Table~\ref{tab:Table_2} we show
the Coulomb excitation cross sections of the first excited state
in each nucleus as a function of the bombarding energy per
nucleon. The cross sections are given in milibarns. The numbers
inside parenthesis and brackets were obtained with pure
non-relativistic and relativistic calculations respectively. The
minimum impact parameter is chosen so that the distance of closest
approach corresponds to the sum of the nuclear radii in a
collision following a Rutherford trajectory. One observes that at
10 MeV/nucleon the relativistic corrections are important only at
the level of 1\%. At 500 MeV/nucleon, the correct treatment of the
recoil corrections (included in the equations~(\ref{IEl})
and~(\ref{IMl})) is relevant on the level of 1\%. Thus the
non-relativistic treatment of Coulomb excitation~\cite{WB65} can
be safely used for energies below about 10 MeV/nucleon and the
relativistic treatment with a straight-line trajectory~\cite{WA79}
is adequate above about 500 MeV/nucleon. However at energies
around 50 to 100 MeV/nucleon, accelerator energies common to most
radioactive beam facilities (MSU, RIKEN, GSI, GANIL), it is very
important to use a correct treatment of recoil and relativistic
effects, both kinematically and dynamically. At these energies,
the corrections can add up to 50\%. These effects were also shown
in Ref.~\cite{AB89} for the case of excitation of giant resonances
in collisions at intermediate energies. As shown here, they are
also relevant for the low-lying excited states.

As another example, we calculate the Coulomb excitation cross sections of
$^{11}$Be projectiles on lead targets. $^{11}$Be is a one neutron halo-nucleus
with one excited bound state ($\frac{1}{2}^{-}$ at 320 keV). Its $\frac{1}%
{2}^{+}$ ground state is strongly coupled to the excited state with the
strongest E1 transition observed between bound nuclear states. The B(E1) value
for this transition is 0.116 e$^{2}$ fm$^{2}$~\cite{Mil83}. The one-neutron
separation energy is only 506 keV and the coupling to the continuum has to be
included in an accurate calculation. However the influence of higher-order
effects in the Coulomb excitation of the excited state in intermediate energy
collisions was shown in Refs.~\cite{BCH95,BT95,SKY96} to be less than 7\%. We
thus neglect these effects here.

\begin{figure}[t]
\begin{center}
\includegraphics[
height=3.1747in, width=3.192in ]{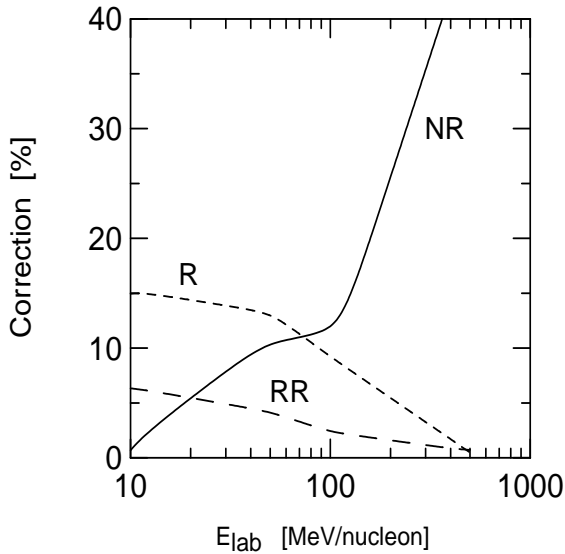}
\end{center}
\caption{ Eq. (\ref{exD}) for the excitation of the 0.89 MeV state
in $^{40}$S + $^{197}$Au collisions as a function of the
bombarding energy. The solid line corresponds to the use of the
non-relativistic integrals Eq.~(\ref{Obtnr}) compared to the exact
calculation using Eq. (\ref{IEl}). The same is plotted for the
other two cases: (R) with the relativistic Eq. (\ref{geez1}), and
(RR) with the relativistic Eq. (\ref{geez1}) using the recoil
correction Eq. (\ref{reccorr}).}
\label{fgrc}%
\end{figure}

In Ref. \cite{WA79} a recoil correction for the theory of
relativistic Coulomb excitation was proposed. It was shown that
one can use the equations for relativistic Coulomb excitation and
obtain reasonable results for collisions at low energies if one
replaces the impact parameter $b$ by
\begin{equation}
b' = b+{\pi \over 2} a \ . \label{reccorr}
\end{equation}
The advantage of this approximation is that one can use the
analytical formulas for relativistic Coulomb excitation (e.g.,
Eqs. \ref{geez1}-\ref{geez2}) and easily include the recoil
correction Eq.~(\ref{reccorr}).  We define the percent deviation
\begin{equation}
\Delta_i = {(\sigma_{\rm exact} - \sigma_i) \over \sigma_{\rm
exact}} \label{exD}
\end{equation}
where $\sigma_i$ is the cross section obtained with the
relativistic ($i=R$), non-relativistic ($i=NR$), and ($i=RR$) with
the relativistic equations for Coulomb excitation
Eq.~(\ref{geez1}) but with the recoil correction
Eq.~(\ref{reccorr}), respectively. Figure \ref{fgrc} is a plot of
Eq. (\ref{exD}) for the excitation of the 0.89 MeV state in
$^{40}$S + $^{197}$Au collisions as a function of the bombarding
energy. One observes that the largest discrepancy is obtained by
using the non-relativistic equations (NR) for the Coulomb
excitation cross sections at high energies. The relativistic
analytical equations (R) also do not do a good job at low
energies, as expected. But the relativistic equations with the
recoil correction of Eq. (\ref{reccorr}) improve considerably the
agreement with the exact calculation. At 10 MeV/nucleon the
deviation from the exact calculation amounts to 6\% for the case
shown in Fig.~\ref{fgrc}. However, the deviation of the RR
treatment tends to increase for cases where higher nuclear
excitation energies are involved \cite{AB89}. The cross section
for the excitation of low energy states is mainly due to
collisions with large impact parameters for which recoil
corrections are not relevant. For high lying states, e.g. giant
resonances, only the smaller impact parameters are effective in
the excitation process. Therefore, in this situation, the correct
treatment of recoil effects is more relevant.

For $^{11}$Be projectiles on lead targets at 50 MeV/nucleon the Coulomb
excitation cross sections of the excited state in $^{11}$Be are given by 311
mb, 305 mb and 398 mb for non-relativistic, exact, and relativistic
calculations, respectively. At 100 MeV/nucleon the same calculations lead to
159 mb, 185 mb and 225 mb, respectively. Thus, the same trend as in the
results of Table~\ref{tab:Table_2} is also observed for E1 excitations of
low-lying states.

Experiments of Coulomb excitation of $^{11}$Be projectiles have been performed
at GANIL (43 MeV/nucleon)~\cite{Ann95}, at RIKEN (64 MeV/nucleon)~\cite{Nak97}
and at MSU (57-60 MeV/nucleon)~\cite{Fau97}. The extracted values of
$B(E1)\sim0.05$ e$^{2}$ fm$^{2}$ in the GANIL experiment is in disagreement
with the lifetime experiment of Ref.~\cite{Mil83} and could not be explained
by higher-order effects in Coulomb excitation at intermediate energy
collisions~\cite{BCH95,BT95,SKY96}. However, the deduced values of
$B(E1)\sim0.1$ e$^{2}$ fm$^{2}$ in the RIKEN and MSU experiments are in good
agreement with the lifetime measurement~\cite{Mil83} and with the theoretical
cross sections of Coulomb excitation. 

\begin{table}[th]
\begin{center}%
\begin{tabular}
[c]{|c|c|c|c|c|c|c|}\hline
Nucleus & E$_{x}$ & B(E2) & 10 MeV/A & 50 MeV/A & 100 MeV/A & 500 MeV/A\\
& [MeV] & [e$^{2}$fm$^{4}$] & $\sigma_{C}$ [mb] & $\sigma_{C}$ [mb] &
$\sigma_{C}$ [mb] & $\sigma_{C}$ [mb]\\\hline
$^{38}$S & 1.29 & 235 & (492) 500 [651] & (80.9) 91.7 [117] & (40.5) 50.1
[57.1] & (9.8) 16.2 [16.3]\\\hline
$^{40}$S & 0.89 & 334 & (877) 883 [1015] & (145.3) 162 [183] & (76.1) 85.5
[93.4] & (9.5) 20.9 [21.]\\\hline
$^{42}$S & 0.89 & 397 & (903) 908 [1235] & (142.7) 158 [175] & (65.1) 80.1
[89.4] & (9.9) 23.2 [23.4]\\\hline
$^{44}$Ar & 1.14 & 345 & (747) 752 [985] & (133) 141 [164] & (63.3) 71.7
[80.5] & (8.6) 17.5 [17.6]\\\hline
$^{46}$Ar & 1.55 & 196 & (404) 408 [521] & (65.8) 74.4 [88.5] & (30.2) 37.4
[41.7] & (5.72)\ 10.8 [11]\\\hline
\end{tabular}
\end{center}
\caption{ Coulomb excitation cross sections of the first excited state in
$^{38,40,42}$S and $^{44,46}$Ar projectiles at 10, 50 100 and 500 MeV/nucleon
incident on gold targets. The numbers inside parenthesis and brackets were
obtained with pure non-relativistic and straight-line relativistic
calculations, respectively. The numbers at the center are obtained with the
full integration of equations~(\ref{IEl}) and~(\ref{IMl}).}%
\label{tab:Table_2}%
\end{table}

\begin{figure}[ptb]
\begin{center}
\includegraphics[
height=3.039in,
width=2.9706in
]
{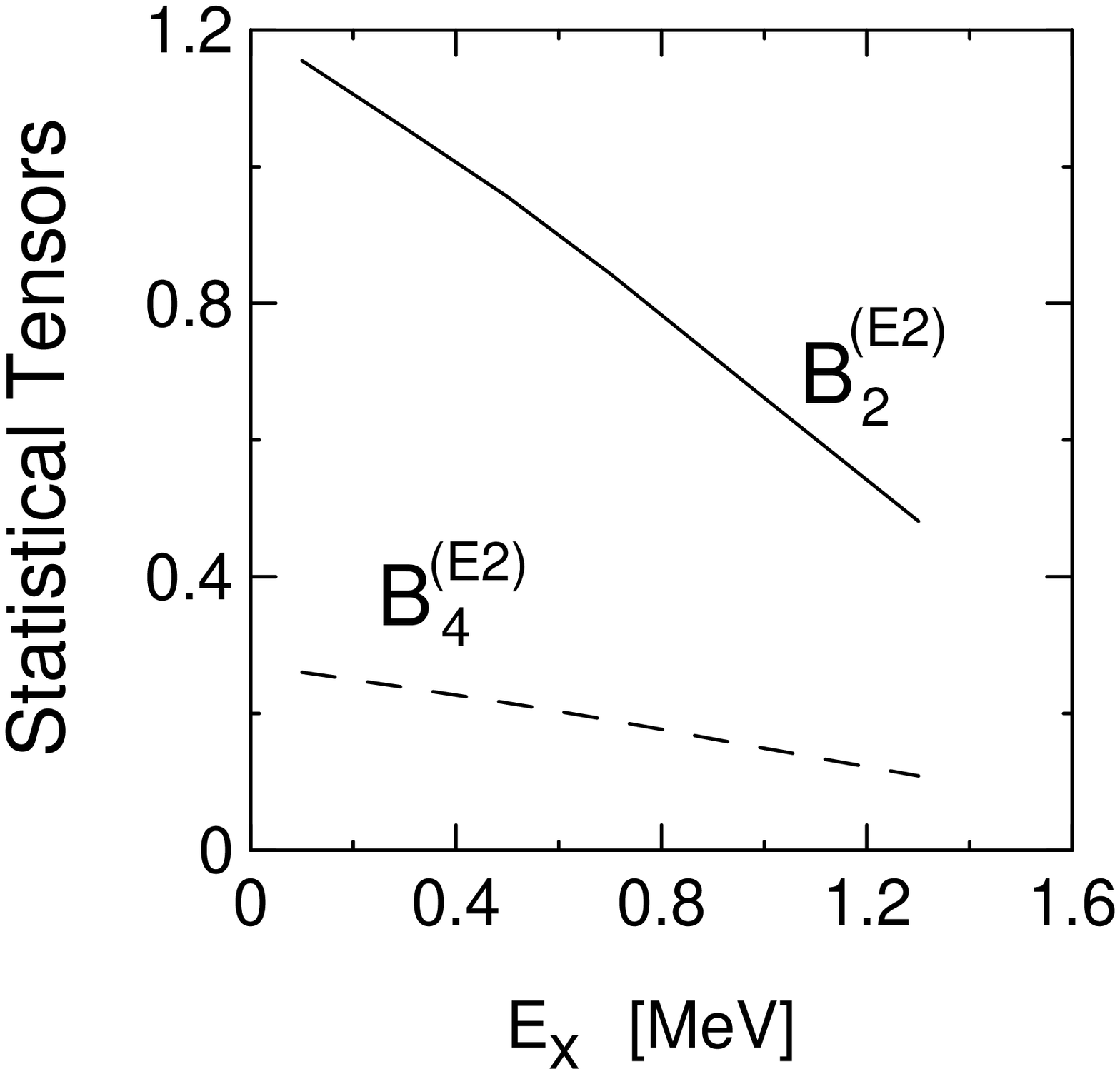}
\end{center}
\caption{ Statistical tensors $B_{2}^{(E2)}$ and $B_{4}^{(E2)}$ as a function
of the excitation energy of the lowest state in $^{38}$S projectiles incident
on gold targets. The energy of the state is varied artificially. }%
\label{a2}%
\end{figure}

We now study the effects of retardation in the angular distributions of
gamma-ray decaying from Coulomb excited states. We first test the range of
validity of the approximation in Eq.~(\ref{ae1ae2}). For this purpose we
artificially vary the energy of the first excited 2$^{+}$ state in $^{38}$S.
This would simulate what happens in the case of a nucleus with very low-lying
excited states, or very high bombarding energies. As we see from
figure~\ref{a2}, the statistical tensors, $B^{(E2)}_{k}$, asymptotically
attain the values $B^{(E2)}_{2}=1.19$ and $B^{(E2)}_{4}=0.267$ according to
the limits in Eq.~(\ref{ae1ae2}) and the definition of $B^{\lambda}_{k}$ in
Eq.~(\ref{Bkl}), since $F_{2} = -0.5976$ and $F_{4} = -1.0690$. The
convergence to these asymptotic values increases with the bombarding energies,
as expected from the conditions which lead to validity of $E_{x}b/\gamma\hbar
c\ll1$, for the lowest impact parameters $b$, which are the most relevant for
the Coulomb excitation process. At 1 GeV/nucleon this condition is easily met
for states of the order of 1 MeV. 

\begin{table}[h]
\begin{center}%
\begin{tabular}
[c]{|c|c|c|c|}\hline
& ~~~NR~~~ & ~~~Exact~~~ & ~~~R~~~\\\hline
$B_{2}^{E2}~~~$ & 0.95 & 1.03 & 1.11\\\hline
$B_{4}^{E2}~~~$ & 0.183 & 0.192 & 0.207\\\hline
\end{tabular}
\end{center}
\caption{ Statistical coefficients entering equation~(\ref{eq:ADNorm}) for
$^{38}$S projectiles at 100 MeV/nucleon incident on gold targets.}%
\label{tab:Table_3}%
\end{table}

Finally, we show in Table~\ref{tab:Table_3} the statistical tensors
$B_{k}^{E2}$ in equation~(\ref{eq:ADNorm}) for $^{38}$S projectiles at 100
MeV/nucleon incident on gold targets and scattering to all kinematically
allowed angles with closest approach distance larger than the sum of the
nuclear radii. NR (R) denotes the non-relativistic (relativistic) values. We
notice that the statistical tensors are not as much influenced by the
retardation and recoil corrections as in the case of the cross sections. The
reason is that the statistical tensors involve ratios of the integral of the
orbital integrals. These ratios tend to wash out the corrections in the
orbital integrals due to relativistic effects. On the other hand, for
scattering to a specific angle the corrections can be larger because the
$\alpha_{k\kappa}$ in Eq.~(\ref{ADgen}) are determined largely by geometry and
hence they can be sensitive to both relativistic distortions of the orbit and
recoil effects causing non straight-line trajectories.

\section{Conclusions}

We have extended the study of Ref.~\cite{AB89} to include retardation effects
in the Coulomb excitation of low-lying states in collisions of rare isotopes
at intermediate energies ($E_{lab}\sim100$ MeV/nucleon). In particular, we
have studied the effects of retardation and recoil in the orbital integrals
entering the calculation of Coulomb excitation amplitudes. We have shown that
the non-relativistic and relativistic theories of Coulomb excitation are
reproduced in the appropriate energy regime. We have also shown that at
intermediate energies corrections to the low- or high-energy theories of
Coulomb excitation are as large as 20\%.

We have studied the excitation of the first excited states in $^{11}$Be,
$^{38,40,42}$S and $^{44,46}$Ar projectiles incident on gold and lead targets.
It is clear from the results that retardation corrections are of the order of
10\%-20\% at bombarding energies around 50-100 MeV/nucleon. Therefore, they
must be accounted for in order to correctly analyze the cross sections and
angular distributions of decaying gamma-rays in experiments at radioactive
beam facilities running at intermediate energies.

Another important consequence of our study is that retardation effects must
also be included in calculations of higher-order effects (e.g.,
coupled-channels calculations), common in the Coulomb breakup of halo
nuclei~\cite{Iek93}. Work in this direction is in progress.

\section*{Acknowledgments}

This research was supported in part by the U.S. National Science Foundation
under Grants No. PHY00-7091, PHY99-83810 and PHY00-70818.

\end{document}